%% file: smcc_ph.tex
\documentclass[aps,prc,twocolumn,showpacs,floatfix,nofootinbib,preprintnumbers,superscriptaddress,amsmath,amssymb,nofootinbib,longbibliography]{revtex4-1}

\usepackage{graphicx}
\usepackage{epsfig}
\usepackage{bm}
\usepackage{color}
\usepackage{float}
\usepackage{dcolumn}
\usepackage{multirow} 
\usepackage{hyperref}
\usepackage{dcolumn}
\usepackage[colorinlistoftodos,prependcaption]{todonotes}

\graphicspath{{.}{figs/}}

\newcommand{\beq}{\begin{equation}}
\newcommand{\eeq}{\end{equation}}
\newcommand{\beqa}{\begin{eqnarray}}
\newcommand{\eeqa}{\end{eqnarray}}

\newcommand{\dbar}[1]{\overline{\overline{\mathcal{#1}}}}
\newcommand{\sbar}[1]{\overline{\mathcal{#1}}}

\newcommand{\ep}[3]{{#1}^{#2}_{#3}}
\begin{document}

\title{Effective shell-model interaction for nuclei southeast of $^{100}$Sn }
\thanks{This manuscript has been authored in part by UT-Battelle, LLC, under contract DE-AC05-00OR22725 with the US Department of Energy (DOE). The US government retains and the publisher, by accepting the article for publication, acknowledges that the US government retains a nonexclusive, paid-up, irrevocable, worldwide license to publish or reproduce the published form of this manuscript, or allow others to do so, for US government purposes. DOE will provide public access to these results of federally sponsored research in accordance with the DOE Public Access Plan (http://energy.gov/downloads/doe-public-access-plan).}

\author{Z.~H.~Sun} 
\affiliation{Department of Physics and Astronomy, University of Tennessee, Knoxville, TN 37996, USA} 
\affiliation{Physics Division, Oak Ridge National Laboratory, Oak Ridge, TN 37831, USA}

\author{G.~Hagen} 
\affiliation{Physics Division, Oak Ridge National Laboratory, Oak Ridge, TN 37831, USA} 
\affiliation{Department of Physics and Astronomy, University of Tennessee, Knoxville, TN 37996, USA}

\author{G.~R.~Jansen} 
\affiliation{National Center for Computational Sciences, Oak Ridge National Laboratory, Oak Ridge, TN 37831, USA} 
\affiliation{Physics Division, Oak Ridge National Laboratory, Oak Ridge, TN 37831, USA}

\author{T.~Papenbrock} 
\affiliation{Department of Physics and Astronomy, University of Tennessee, Knoxville, TN 37996, USA} 
\affiliation{Physics Division, Oak Ridge National Laboratory, Oak Ridge, TN 37831, USA}

\begin{abstract} 
We construct an effective shell-model interaction for the valence space spanned by single-particle neutron  and single-hole proton states in $^{100}$Sn. Starting from chiral nucleon-nucleon and three-nucleon forces and single-reference coupled-cluster theory for $^{100}$Sn we apply a second similarity transformation that decouples the valence space. The particle-particle components of the resulting effective interaction can be used in shell model calculations for neutron deficient tin isotopes. The hole-hole interaction can be used to calculate the $N = 50$ isotones south of $^{100}$Sn, and the full particle-hole interaction describes nuclei in the region southeast of $^{100}$Sn. We compute low-lying excited states in selected nuclei southeast of $^{100}$Sn, and find reasonable agreement with data. The presented techniques can also be applied to construct effective shell-model interactions for other regions of the nuclear chart.
\end{abstract}

\maketitle


\input{introduction}

\input{method}

\input{results}

\begin{acknowledgments}
This material is based upon work supported by the U.S. Department of Energy, Office of Science, Office of Nuclear Physics under Award
  Numbers DEFG02-96ER40963 (University of Tennessee), DE-SC0008499
  (SciDAC-3 NUCLEI), DE-SC0018223 (SciDAC-4 NUCLEI), DE-SC0015376
  (Double-Beta Decay Topical Collaboration), and the Field Work
  Proposals ERKBP57 and ERKBP72 at Oak Ridge National Laboratory
  (ORNL).  Computer time was provided by the Innovative and Novel
  Computational Impact on Theory and Experiment (INCITE) program. This
  research used resources of the Oak Ridge Leadership Computing
  Facility located at ORNL, which is supported by the Office of
  Science of the Department of Energy under Contract No.
  DE-AC05-00OR22725.
\end{acknowledgments}


%

\end{document}

%% file: introduction.tex
\section{Introduction}
The shell model is the paradigm to understand the structure of atomic nuclei and to compute their properties~\cite{Mayer1955,brown1988,otsuka2001,caurier2005}. In this model, valence nucleons move within the mean-field produced by the inert core and interact via a residual effective interaction. For a given model space, the matrix elements of the effective interaction can be determined  phenomenologically by fit to data~\cite{Brown2001,Brown2006,honma2004}, by tweaking the monopole terms~\cite{poves1981} of a microscopically derived $G$-matrix~\cite{hjorthjensen1995}, from a low-momentum nucleon-nucleon potential~\cite{bogner2003} by choosing a suitable cutoff~\cite{Coraggio2009a,Coraggio2011}, or, more recently, derived without adjustable parameters from nucleon-nucleon and three-nucleon potentials~\cite{bogner2014,jansen2014,jansen2015,stroberg2016,dikmen2015,sun2018}. Several of these approaches requires one to decouple the two-body interaction in a small shell-model space from the (infinite) Hilbert space. Here, similarity transformations play a key role. These can be based on the Lee-Suzuki approaches~\cite{suzuki1980,suzuki1982}, coupled-cluster theory~\cite{kuemmel1978,jansen2014,hagen2014}, or the in-medium similarity renormalization group (IMSRG) ~\cite{tsukiyama2011,tsukiyama2012,hergert2016}.

In this paper, we develop an effective shell-model interaction for nuclei southeast of $^{100}$Sn. This region of the nuclear chart is interesting because its cornerstone -- the doubly-magic $N=Z=50$ nucleus $^{100}$Sn -- is close to the proton dripline, the endpoint of $\alpha$-decays~\cite{seweryniak2006,darby2010}, and exhibits one of the strongest Gamow-Teller matrix elements~\cite{hinke2012,gysbers2019,lubos2019}. Several recent experiments studied the structure of nuclei in this part of the nuclear chart~\cite{herfurth2011,ferrer2014,xu2019,hornung2020,mougeot2021}. Traditionally, shell-model computations of nuclei in this region start from a $^{88}$Sr or $^{90}$Zr core~\cite{faestermann2013,morris2018}. In such an approach, the proton shell is almost full once elements close to tin have been reached, in addition, the computational cost quickly becomes a bottleneck (see e.g. ~\cite{morris2018}), which makes it attractive to compute a particle-hole shell-model interaction starting from $^{100}$Sn. We note that $^{100}$Sn can be computed from scratch using coupled-cluster theory~\cite{morris2018}, and this nucleus is predicted to be a ``better," i.e. more strongly bound, core than $^{88}$Sr~\cite{faestermann2013}.

Recent non-perturbative approaches to effective interactions include methods that are based on many-body wavefunctions calculated with {\it ab-initio} methods such as no-core shell model (NCSM) \cite{Lisetskiy2008,dikmen2015} and coupled-cluster theory \cite{jansen2014,jansen2016}.
Once the wavefunctions of $A_{\text{core}}$+1 and $A_{\text{core}}$+2 are obtained, the effective interaction can be extracted through a Lee-Suzuki transformation~\cite{suzuki1980,Suzuki1994d,Suzuki1982a, Okubo1954a}. 
Effective interactions derived through these approaches decouple the model space from the excluded space on the two- and three-body levels without introducing additional parameters. In principle, these non-perturbative methods also work for multi-shell spaces. However, it is challenging to obtain converged full-space wavefunctions for all allowed quantum numbers defined by the valence space, and for heavy nuclei. In the coupled-cluster implementation, for instance, some observables are sensitive to small contributions from high-lying excited states. 

This computational problem is, to some extent, overcome by the valence-space IMSRG (VS-IMSRG) \cite{Hergert2016c,Hergert2017,Stroberg2017}, and the shell-model coupled-cluster (SMCC)~\cite{sun2018} approaches. Instead of employing the $A$-body wavefunctions in Hilbert space, VS-IMSRG and SMCC are straightforward extensions of IMSRG and coupled-cluster theory, respectively, to open-shell nuclei. Similarity transformations are applied to the normal-ordered Hamiltonian to decouple the model space and these approaches are suitable to calculate medium-heavy nuclei. Recently the VS-IMSRG was successfully applied to compute nuclei in the island of inversion region by decoupling a multi-shell valence space~\cite{miyagi2020}.

In this paper we select the core nearest to the target nuclei and perform a shell model calculation with both particles and holes. The effective interaction is derived through the particle-hole shell-model coupled-cluster approach (ph-SMCC). In ph-SMCC we treat the three-body correlations in a single-reference coupled-cluster approach instead of in the shell-model effective interaction. This is an advantage because the coupled-cluster method is more accurate than SMCC in dealing with the induced three-body correlations.
The ph-SMCC is an extension of SMCC, in which a secondary similarity transformation is applied to the coupled-cluster effective Hamiltonian. The resulting Hamiltonian decouples the excluded space and the valence space, and the latter is spanned by particle and hole states.   Once the Hamiltonian is decoupled, the valence effective interaction consists of particle-particle({\it pp}), hole-hole({\it hh}), and particle-hole({\it ph}) channels. The {\it pp} channel is a conventional shell model effective interaction, and particle-removed nuclei can be calculated with the {\it hh} interaction. The {\it ph} channel can be used to calculate nuclei located in the south-east and north-west of a double magic nucleus, i.e., as protons removed, and neutrons attached, or vice versa. 
The particle-hole decoupling is equivalent to a conventional decoupling with the core in the valence space, and this is similar to the ensemble normal-ordering used in the VS-IMSRG~\cite{Stroberg2017}. The resulted effective particle-hole interaction can be re-normal-ordered with respect to a smaller core to obtain a more conventional {\it pp} interaction. 

The paper is organized as follows. In Section~\ref{sec:PH}, we briefly introduce the single-reference coupled-cluster method and the resulting similarity-transformed Hamiltonian. This is followed by the derivation of the particle-hole decoupling. In Section~\ref{sec:results} we apply the ph-SMCC method to the $^{100}$Sn region using two different chiral nucleon-nucleon and three-nucleon potentials. 
Finally, we summarize our results.

%% file: method.tex
\section{Shell-model coupled-cluster particle and hole interaction }
\label{sec:PH}

\subsection{Shell-model coupled cluster}
The coupled-cluster method~\cite{coester1958,kuemmel1978,dean2004,bartlett2007,binder2013,hagen2014} is useful for ab-initio calculations of medium-mass nuclei~\cite{hagen2016,morris2017}. 
Our coupled-cluster calculations  start from the intrinsic  Hamiltonian,
\begin{equation}
\label{insh}
H=\left(1-\frac1{A}\right)\sum_{i=1}^A\frac{p_i^2}{2m}+ \sum_{i<j=1}^A \left(v_{ij}-\frac{\overrightarrow{p_i}\cdot \overrightarrow{p_j}}{mA} \right)+\sum_{i<j<k}^Av_{ijk}.
\end{equation}
Here $m$ is the mass of nucleon, $p_i$ and $p_j$ are the single-particle momentum, $A$ is the mass number, $v_{ij}$ is the nucleon-nucleon potential and $v_{ijk}$ is the three-body potential defined in the laboratory coordinates. The Hamiltonian is henceforth normal-ordered with respect to a reference state $|\Phi_0\rangle$, e.g. a Hartree-Fock (HF) state or a product state of natural orbitals. We denote the energy expectation value of the reference state as $E_0=\langle\Phi_0|H|\Phi_0\rangle$. 
The normal-ordered Hamiltonian is
\begin{eqnarray}
\label{nmod1}
\mathcal{H}=& & E_0+\sum_{pq}f_{pq}\{p^\dagger q\}+\frac14\sum_{pqrs}V_{pqrs} \{p^\dagger q^\dagger sr\}\\
&+&\frac1{36}\sum_{pqrstu}V_{pqr,stu}\{p^\dag q^\dag r^\dag s t u\}.
\end{eqnarray} 

Here we use $p^\dagger$ as the particle or hole creation operator on state $|\phi_p\rangle$, and $p$ is the annihilation operator. The one-body Fock matrix has elements $f_{pq}$, while $V_{pqrs}$ and $V_{pqr,stu}$ denote the two-body and three-body matrix elements, respectively. To avoid dealing with the three-body diagrams after normal-ordering, Eq.(\ref{nmod1}) is usually truncated at the two-body level and the residual three-body terms are discarded. This approximation is accurate in $^4$He~\cite{hagen2007a}, $^{16}$O~\cite{roth2012}, and nuclear matter~\cite{hagen2016b} (albeit only for three-nucleon forces with nonlocal regulators). 

Coupled cluster theory is based on the exponential ansatz
\begin{equation}
	|\Phi\rangle=e^{T}|\Phi_0\rangle , 
\end{equation}
where $T$ is the cluster operator 
\begin{equation}\label{tamp}
	T=T_1+T_2+T_3+\cdots ,
\end{equation}
that introduces 1-particle--1-hole ($1p$--$1h$), $2p$--$2h$, $3p$--$3h$, ...$Ap$--$Ah$ excitations.
The $np$--$nh$ excitation operator is
\begin{equation}
	T_n=\frac1{(n!)^2}\sum_{\substack{i_1,i_2,\cdots, i_n \\ a_1,a_2,\cdots, a_n}} t_{i_1,i_2,\cdots, i_n}^{a_1,a_2,\cdots, a_n}\{a_1^\dag a_2^\dag \cdots a_n^\dag i_1i_2\cdots i_n\}.
\end{equation}
The Schr\"odinger equation is then written as
\begin{equation}\label{schroeq}
	\mathcal{H}e^T|\Phi_0\rangle =Ee^T|\Phi_0\rangle.
\end{equation}
Left multiplication with $e^{-T}$ on both sides of Eq.(\ref{schroeq}) yields
\begin{equation}\label{schroeq2}
	e^{-T}\mathcal{H}e^T|\Phi_0\rangle =E|\Phi_0\rangle .
\end{equation}
Equation~(\ref{schroeq2}) indicates that the reference state $|\Phi_0\rangle$ is an eigenstate of the similarity-transformed Hamiltonian 
\begin{equation}
\overline{\mathcal{H}}\equiv  e^{-T}\mathcal{H}e^T \ . 
\end{equation}
In other words, $\overline{\mathcal{H}}$ generates no $ph$ excitations of the reference, and 
\begin{equation}\label{deceq}
	\langle \Phi_{i_1i_2\cdots i_n}^{a_1 a_2 \cdots a_n}|\sbar{H}|\Phi_0\rangle =0 \ .
\end{equation}
The $T$ amplitudes fulfill Eq.(\ref{deceq}) and thereby decouple the reference state from all excited states. The similarity transformed Hamiltonian can be expanded as
 \begin{eqnarray}
\sbar{H}& = &E_{CC}+\sum_{pq}\sbar{H}_{pq}\{p^\dagger q\}+\sum_{pqrs} \sbar{H}_{pqrs}\{p^\dagger q^\dagger sr\}\nonumber \notag \\
&\quad&+\sum_{pqrstu} \sbar{H}_{pqrstu}\{p^\dagger q^\dagger r^\dagger uts\}+\cdots , \label{hccsd}
\end{eqnarray}
and contains -- when $T$ is not truncated --  $A$-body operators, even when the normal-ordered Hamiltonian $\mathcal{H}$ has lower rank. Here, $E_{CC}=\langle \Phi_0|\sbar{H}|\Phi_0\rangle$ is the correlation energy.
In practice, the cluster operator in Eq.~(\ref{tamp}) is truncated to avoid the exponential computational expense of full configuration mixing.  Truncation  to $T=T_1+T_2$ yields the coupled-cluster singles and doubles (CCSD) approximation. 

Single-reference coupled-cluster theory is most efficient  to calculate closed-shell nuclei. Open-shell systems can be computed by starting from a deformed, symmetry-breaking reference state~\cite{novario2020a}, and  the broken symmetry needs to be restored through angular momentum projection~\cite {Qiu2017,Yuan2000}. Another approach to open-shell nuclei employs  equation-of-motion methods (EOM-CC)\cite{jansen2011,jansen2012,jansen2015,ekstrom2015a};  these are restricted to neighbors of closed-shell nuclei

\begin{figure}[ht]
  \includegraphics[width=\linewidth]{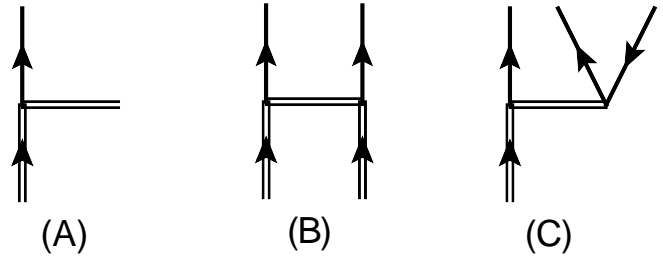}
  \caption{Diagrammatic representation of the decoupling generator $S_\text{pp}$. The horizontal line is the $S_\text{pp}$ operator, with the particles indicated by incoming and outgoing arrow lines. The model-space particles indicated by incoming double lines.}
   \label{fig.qp}
\end{figure}

The SMCC method is an alternative approach to treat the open-shell problem with coupled-cluster methods by creating an effective interaction to be used in shell-model calculations. In such a calculation, the effective interaction is defined in a model space spanned by one or two major shells, whereas $\sbar{H}$ is defined in the full space defined by several shells in the harmonic-oscillator basis.
To construct a shell-model effective interaction, the Hilbert space is split into the model space $P$, and its complement $Q$, where
\begin{equation}
	P+Q=\mathbf{1}.
\end{equation}
The task is then to decouple the $P$ space from the $Q$ space for the Hamiltonian. 
To accomplish this, a secondary similarity transformation is applied to the coupled-cluster Hamiltonian,
\begin{equation}\label{dec2}
	\dbar{H}=e^{-S}\sbar{H}e^S,
\end{equation}
and the decoupling between $P$ and $Q$ space requires
\begin{equation}\label{adec}
	Q\dbar{H}P=0.
\end{equation}
Equation~(\ref{adec}) is an $A$-body equation and $S$ is also an $A$-body operator. In what  follows, we use $S_\text{pp}$ to represent the $S$ operator in particle-particle decoupling, $S_\text{hh}$ for that in the hole-hole decoupling, $S_\text{hp}$ for particle and hole decoupling and $S$ with no subscript for any combination of them. On the two-body level, $S_\text{pp}$ has the form
\begin{equation}
\begin{aligned}
S_\text{pp} ={} & \sum_{av_1}s^a_{v_1}\{ a^\dag v_1\}+\sum_{abv_1v_2}s^{ab}_{v_1v_2}\{a^\dag b^\dag v_1 v_2\} \\
     +&\sum_{abhv}s^{ab}_{hv}\{a^\dag b^\dag h v\}+\cdots.
\end{aligned}
\label{Seq}
\end{equation}
Here we used $v$ to represent valence state, while $a,b,c,\cdots$ and $i,j,k,\cdots$ denote particle and hole states, respectively. The first term in Eq.~(\ref{Seq}) represents a valence particle $v_1$ coupled to a state outside the model space $a$ through a one-body operator [see diagram (a) in Fig.~\ref{fig.qp}], the second term couples two particles inside the model space to a state which has at least one particle outside the model space [see diagram (b) in Fig.~\ref{fig.qp})], and the third term couples the valence state to particle-hole configurations [see diagram (c) in Fig.~\ref{fig.qp}]. One can write down three-body terms (and operators of higher rank) in a similar manner.

In this paper, we truncate Eqs.~(\ref{adec}) and (\ref{Seq}) at the two-body level and refer to this as the SMCC(2) approximation. The two-body level decoupling condition~(\ref{adec}) becomes
 \begin{eqnarray}\label{hhdec}
\langle a| \dbar{H} | v_1\rangle &=& 0 \label{deca}\\
\langle a b| \dbar{H} | v_1 v_2\rangle &=& 0 \label{decb}\\
\langle a b| \dbar{H} | i v_1\rangle &=& 0\label{decc}.
\end{eqnarray}

Once decoupled, the effective Hamiltonian is
\begin{equation}
	H_{\text{eff}}=P\dbar{H}P,
\end{equation}
which is defined only inside the model space.
In principle, the Hamiltonian $H_{\text{eff}}$ reproduces a subset of the eigenvalues of $\sbar{H}$. In practice, however, the truncation of many-body terms leads to a discrepancy of eigenvalues between $H_{\text{eff}}$ and $\dbar{H}$, because the truncation of many-body terms breaks the similarity of the transformation. The quality of the effective interaction thus depends on whether the neglected many-body terms are small compared to the retained two-body matrix elements~\cite{sun2018}.

The evaluation of $\dbar{H}$ is nontrivial due to the properties of $S$. In single-reference coupled-cluster theory, the operator $T$ only consists of excitation operators, and we have
\begin{equation}
	\left[T_m,T_n\right]=0 ,
\end{equation} 
i.e., different excitations commute. Thus, $H$ can only contract with $T$ from the right side, and Eq.~(\ref{schroeq2}) terminates exactly on the fourth nested commutator in CCSD for two-body Hamiltonians. In contrast, the operator $S$ in SMCC consists of both excitation and de-excitation operators, and these do not commute with each other, i.e. 
  \begin{equation}
  	[S_m,S_n]\neq 0.
  \end{equation}
Consequently, $\sbar{H}$ can contract with $S$ from both the left and the right side and the evaluation of $\dbar{H}$ does not terminate within the  Baker–-Campbell--Hausdorff (BCH) expansion,
      \begin{equation}\label{smccbch}
	\dbar{H}=\sbar{H}+[\sbar{H},S]+\frac{1}{2!}[[\sbar{H},S],S]+\cdots \ .
  	 \end{equation}
To perform the similarity transformation we adopt an approach that is similar to the Magnus method from the IMSRG~\cite{Morris2015}. Here, $S$ is solved through an integral equation~\cite{sun2018}
  \begin{equation}
  S=\int \eta(\dbar{H}) dt \ , 
  \end{equation}
where $t$ is the flow parameter, and $\eta$ is a function of the off-diagonal part of $\dbar{H}$. When performing the integration,  
$\dbar{H}$ develops more than two-body terms. The induced three-body force can link the valence space and the excluded space via an off-diagonal three-body term $S_{3b}$. The diagonal three-body terms also contribute to the BCH expansion through  $[H_{\text{3b}},S_{\text{2b}}]_{\text{2b}}$.
The explicit evaluation of the three-body terms is challenging. We follow the two-body normal-ordered IMSRG calculations and truncate each commutator at the  two-body level, and nested commutators are computed recursively via
  	  	\begin{equation}\label{commt}
  		D^{n}=[D^{n-1},S]_{\text{2b}} \ .
  	\end{equation}
  	Here, $D^{(0)}=\sbar{H}$. This allows us to re-write Eq.~(\ref{smccbch}) as
  	\begin{equation}
  		\dbar{H}=\sum_{n=0}\frac1{n!}D^{(n)} \ .
  	\end{equation}

To estimate the contribution of the neglected terms, the leading-order from $[[D,S_\text{2b}]_\text{3b},S_{\text{2b}}]_\text{2b}$ was studied in Ref.~\cite{sun2018}, and  this commutator yields a small contribution compared to the leading-order contribution from $S_\text{3b}$. 
Based on these results, we also truncate operators in this work at the two-body level and add the leading-order contribution from $S_\textbf{3b}$  perturbatively to $\dbar{H}$. 
   
The main task is then to calculate the contraction $[D^{(n)},S]$. 
We find
  \begin{eqnarray}
  	\ep{D^{(n+1)}}{ia}{bc}&=& (1-P_{bc})\left(\sum_e\ep{D^{(n)}}{ia}{be}\ep{S}{e}{c}+\sum_{je}\ep{D^{(n)}}{ij}{be}\ep{S}{ea}{jc}\right)\nonumber\\
  	&-&\sum_{e}\left(\ep{D^{(n)}}{ie}{bc}\ep{S}{a}{e}-\ep{D^{(n)}}{i}{e}\ep{S}{ea}{bc}\right)\nonumber \\
  	&+&\frac12 \sum_{ef}\ep{D^{(n)}}{ia}{ef}\ep{S}{ef}{bc} \ , \\
  	 \ep{D^{(n+1)}}{ij}{ka}&=&\frac12 \sum_{ef}\ep{D^{(n)}}{ij}{ef}\ep{S}{ef}{ka}+\sum_{e}\ep{D^{(n)}}{ij}{ke}\ep{S}{e}{a} \ , \\
  	 \ep{D^{(n+1)}}{ia}{jb}&=&\sum_{e}\left( \ep{D^{(n)}}{ia}{je}\ep{S}{e}{b}-\ep{D^{(n)}}{ie}{jb}\ep{S}{a}{e}+\ep{D^{(n)}}{i}{e}\ep{S}{ea}{jb}\right) \nonumber \\
  	 &+&\frac12 \sum_{ef}\ep{D^{(n)}}{ia}{ef}\ep{S}{ef}{jb}+\sum_{ke}\ep{D^{(n)}}{ik}{je}\ep{S}{ea}{kb} \ , \\
  	  \ep{D^{(n+1)}}{ab}{cd}&=&\frac12 \sum_{ef}\left( \ep{D^{(n)}}{ab}{ef}\ep{S}{ef}{cd}-\ep{S}{ab}{ef}\ep{D^{(n)}}{ef}{cd}\right )\nonumber \\
  	  			&+&\sum_{e}(1-P_{cd})\left( \ep{D^{(n)}}{ab}{ce}\ep{S}{e}{d}-\ep{S}{ab}{ce}\ep{D^{(n)}}{e}{d}\right)\nonumber \\ 
  	  			&-&\sum_{e}(1-P_{ab})\left(\ep{D^{(n)}}{eb}{cd}\ep{S}{a}{e}-\ep{S}{eb}{cd}\ep{D^{(n)}}{a}{e} \right )\nonumber \\
  	  			&-&\sum_k(1-P_{cd})\ep{D^{(n)}}{k}{c}\ep{S}{ab}{kd}\nonumber \\
  	  			&+&\sum_{je}(1-P_{ab})(1-P_{cd})\ep{D^{(n)}}{ja}{ec}\ep{S}{eb}{jd} \ ,\\
  	  			\ep{D^{(n+1)}}{ij}{kl}&=& 0\ ,  \label{ppzero}
  	 \end{eqnarray}
  	 and
  	 \begin{eqnarray}
  	  \ep{D^{(n+1)}}{ab}{ic} &=& \frac12\sum_{ef} \left(\ep{D^{(n)}}{ab}{ef}\ep{S}{ef}{ic}-\ep{S}{ab}{ef}\ep{D^{(n)}}{ef}{ic}\right)\nonumber\\
  	   	  &+&\sum_{e}\left( \ep{D^{(n)}}{ab}{ie}\ep{S}{e}{c}-\ep{D^{(n)}}{e}{c}\ep{S}{ab}{ie}\right) \nonumber \\
  	   	  &+&\sum_{e}(1-P_{ab})\left(\ep{D^{(n)}}{b}{e}\ep{S}{ae}{ic}-\ep{S}{b}{e}\ep{D^{(n)}}{ae}{ic}\right)\nonumber \\
  	   	  &-&\sum_k \ep{S}{ab}{kc}\ep{D^{(n)}}{k}{i}\nonumber \\
  	   	  &-&\sum_{je}(1-P_{ab})\ep{D^{(n)}}{ja}{ie}\ep{S}{eb}{jc} \ , \\
  	   \ep{D^{(n+1)}}{ia}{jk} &=&-\sum_{e}\ep{D^{(n)}}{ie}{jk}\ep{S}{a}{e} \ , \\
  	  \ep{D^{(n+1)}}{ij}{ab}&=& \frac12 \sum_{ef}\ep{D^{(n)}}{ij}{ef}\ep{S}{ef}{ab}\nonumber \\
  	  &+&\sum_{e}(1-P_{ab})\ep{D^{(n)}}{ij}{ae}\ep{S}{e}{b} \ .
  \end{eqnarray}
We make an initial guess of the $S$ operator, compute the operator $\dbar{H}$ and update  $S$ accordingly. By repeating this procedure we integrate out the off-diagonal pieces of $\sbar{H}$ until the decoupling condition is fulfilled.

\subsection{Particle-hole shell model coupled cluster}
\begin{figure}[hbt]
  \includegraphics[width=0.48\textwidth]{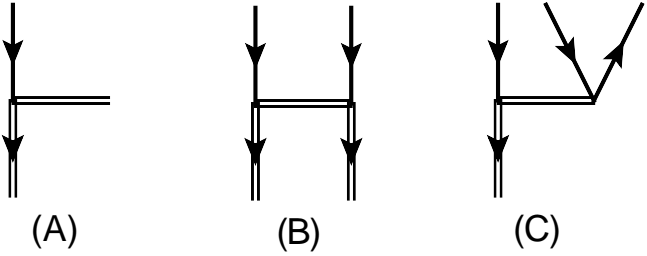}
  \caption{Diagrammatic representation of $S_\text{hh}$. The horizontal line is the $S_\text{hh}$ operator, with the particles and holes indicated by incoming and outgoing arrow lines. The model-space holes indicated by outgoing double lines}
  \label{fig.shh}
\end{figure}

Before moving to the particle-hole decoupling, we derive the effective hole-hole interaction. A hole-hole decoupling yields an effective interaction for valence holes. On the two-body level, the operator $S$ for hole-hole decoupling is
\begin{equation}\label{shh}
	S_{\text{hh}}=\sum_{iv_1}s^{v_1}_i\{v_1^\dag i\}+\sum_{v_1v_2ij}s^{v_1 v_2}_{ij}\{v_1^\dag v_2^\dag ij\}
	+\sum_{ijv_1a}s^{v_1a}_{ij}\{v_1^\dag a^\dag i j\} \ .
\end{equation}
Here, $v_i $ denotes a valence-hole state. The first two terms link  valence holes and  holes outside the model space through a one- and two-body operator, and the last term is the single hole coupled to the particle-hole excitation configurations. Diagrams representing Eq.~(\ref{shh}) are shown in Fig.~\ref{fig.shh}.

Similar to the particle-particle decoupling,
$S_{\text{hh}}$ is determined through solving the decoupling equation~(\ref{adec}). The two-body level decoupling equation for the hole-hole effective interaction is
\begin{eqnarray}
\langle v_1| \dbar{H} | i\rangle &=& 0 \label{hhdec1}\\
\langle v_1 v_2| \dbar{H} | ij\rangle &=& 0 \label{hhdec2}\\
\langle v_1 a| \dbar{H} | i j\rangle &=& 0.\label{hhdec3}.
\end{eqnarray}
Once decoupled, the hole-hole sector of $\dbar{H}$ should reproduce a subset of eigenvalues of particle removed systems.

Using the two-body level $S$ for the  hole-hole decoupling in Eq.~(\ref{shh}), the single commutator yields
\begin{eqnarray}
	\ep{D^{(n+1)}}{ia}{bc}&=&\frac12 \sum_{kl}\ep{D^{(n)}}{kl}{bc}\ep{S}{ia}{kl}-\sum_{k}\ep{D^{(n)}}{ka}{bc}\ep{S}{i}{k} \ , \\
  	 \ep{D^{(n+1)}}{ij}{ka}&=&-\sum_l(1-P_{ij})\ep{D^{(n)}}{lj}{ka}\ep{S}{i}{l}\nonumber \\
  	 &+&\sum_{bl}(1-P_{ij})\ep{D^{(n)}}{lj}{ba}\ep{S}{ib}{kl} \nonumber\\
  	 &+&\sum_l \left (\ep{D^{(n)}}{ij}{la}\ep{S}{l}{k}-\ep{S}{ij}{kl}\ep{D^{(n)}}{l}{a} \right)\nonumber \\
  	 &+&
  	 \frac12 \sum_{lm}\ep{D^{(n)}}{lm}{ka}\ep{S}{ij}{lm} \ , \\
  	 \ep{D^{(n+1)}}{ia}{jb}&=&\sum_{k}\left( -\ep{D^{(n)}}{ka}{jb}\ep{S}{i}{k}+\ep{D^{(n)}}{ia}{kb}\ep{S}{k}{j}-\ep{D^{(n)}}{k}{b}\ep{S}{ia}{jk}\right)\nonumber \\
  	 &+&
  	 \frac12 \sum_{kl}\ep{D^{(n)}}{kl}{jb}\ep{S}{ia}{kl}+\sum_{ck}\ep{D^{(n)}}{ka}{cb}\ep{S}{ic}{jk} \ , \\
  	 \ep{D^{(n+1)}}{ab}{cd}&=&0 \ , \label{hhzero}
  	 \end{eqnarray}
  	 and
  	 \begin{eqnarray}
  	  \ep{D^{(n+1)}}{ij}{kl}&=& \frac12\sum_{mn}\left( \ep{D^{(n)}}{mn}{kl}\ep{S}{ij}{mn} - \ep{S}{mn}{kl}\ep{D^{(n)}}{ij}{mn}\right) \nonumber\\
  	     &+& \sum_m(1-P_{ij})\left (-\ep{D^{(n)}}{im}{kl}\ep{S}{j}{m}+\ep{S}{im}{kl}\ep{D^{(n)}}{j}{m}\right)\nonumber \\
  	     &+&\sum_m(1-P_{kl})\left( \ep{D^{(n)}}{ij}{km}\ep{S}{m}{l}-\ep{S}{ij}{km}\ep{D}{m}{l} \right ) \nonumber \\
  	     &+& \sum_a (1-P_{ij})\ep{D^{(n)}}{j}{a}\ep{S}{ia}{kl}\nonumber \\
  	     &+&\sum_{am}(1-P_{ij})(1-P_{kl})\ep{D^{(n)}}{im}{ka}\ep{S}{ja}{lm} \ , \\
  	  \ep{D^{(n+1)}}{ab}{ic} &=& \sum_{k}\ep{D^{(n)}}{ab}{kc}\ep{S}{k}{i} \ , \\
  	   \ep{D^{(n+1)}}{ia}{jk} &=&\frac12 \sum_{mn}\left( \ep{D^{(n)}}{mn}{jk}\ep{S}{ia}{mn}- \ep{S}{mn}{jk}\ep{D^{(n)}}{ia}{mn} \right) \nonumber\\
  	    &+&\sum_l\left( -\ep{D^{(n)}}{la}{jk}\ep{S}{i}{l}+\ep{S}{la}{jk}\ep{D^{(n)}}{i}{l} \right)  \nonumber\\
  	    &+& \sum_l(1-P_{jk})\left(\ep{D^{(n)}}{ia}{lk}\ep{S}{l}{j}-\ep{S}{ia}{lk}\ep{D^{(n)}}{l}{j} \right) \nonumber \\
  	    &+&\sum_b \ep{D^{(n)}}{a}{b}\ep{S}{ib}{jk}\nonumber \\
  	    &-& \sum_{cl}(1-P_{jk})\ep{D^{(n)}}{la}{kc}\ep{S}{ic}{jl} \ , \\
  	  \ep{D^{(n+1)}}{ij}{ab}&=&\frac12 \sum_{kl}\ep{D^{(n)}}{kl}{ab}\ep{S}{ij}{kl}\nonumber\\
  	  &-& \sum_{k}(1-P_{ij})\ep{D^{(n)}}{ik}{ab}\ep{S}{j}{k} \ .
\end{eqnarray}

Equations~(\ref{ppzero}) and (\ref{hhzero}) imply that the hole-hole interaction is not affected by a particle-particle decoupling, meanwhile the particle-particle interaction is not changed by a hole-hole decoupling. This is true only in a single commutator, and the interplay between the {\it hh} and {\it pp} decoupling occurs in the nested commutator. This suggests that the {\it pp} interaction is weekly affected by the {\it hh} decoupling, because of the diminishing importance of increasingly nested  commutators. In practical calculations, the effective {\it pp} interaction from a particle-hole decoupling is not identical to that from a {\it pp} decoupling, but as expected they do reproduce the same eigenvalues. 

\begin{figure}[hbt]
  \includegraphics[width=0.45\textwidth]{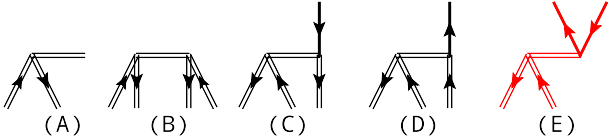}
  \caption{Diagrammatic representation of $S_\text{hp}$. The horizontal line is the $S_\text{hp}$ operator, with the particles indicated by incoming and outgoing arrow lines. The valence space particles and holes are indicated by incoming double lines. (A) represents a particle-hole de-excitation in the valence space, (B) a valence space $2p$--$2h$ de-excitation, (C) a valence-space $2h$--$1p$ coupled to one hole, (D) a valence-space $2p$--$1h$ coupled to one particle, and (E) represent valence-space particle-hole coupled to a particle-hole in the excluded space.}
  \label{fig.shp}
\end{figure}

In the particle-hole decoupling, valence particles and holes need to be treated on an equal footing. The possibility of particle-hole de-excitation in the valence space complicates the decoupling, as each sector of the Hamiltonian need to be decoupled, see Fig.~\ref{fig.shp}. In most shell-model calculations, we do not usually need protons or neutrons to be both particle and hole states simultaneously. In the current work, we restrict ourselves to a model space that contains only neutron particles and proton hole states (or verse vice). The off-diagonal Hamiltonian contains no particle-hole de-excitations due to isospin conservation, and only the last term in Fig.~\ref{fig.shp} contributes. The $S$ operator for the particle-hole decoupling is
\begin{equation}
	S=S_\text{hh}+S_\text{pp}+S_\text{hp}.
\end{equation}
Here, $S_\text{pp}$ and $S_\text{hh}$ have already been discussed above and we will focus on the additional terms. Diagram E in Fig.~\ref{fig.shp} is
\begin{equation}
\label{shpeq}
	S_{\text{hp}}=\sum_{iajv}s^{v_1 a}_{j v_2}\{v_1^\dag a^\dag j v_2\}.
\end{equation}
Here, $S_{\text{hp}}$ couples the particle-hole states between the model space and the excluded space.

Additional terms contributing to the single commutator $[D^{(n)},S_\text{hp}]$ are
\begin{eqnarray}
	\ep{D^{(n+1)}}{ia}{bc}&=& \sum_{kd} (1-P_{bc})\left( \ep{D^{(n)}}{ka}{cd}\ep{S}{id}{kb}- \ep{S}{ka}{cd}\ep{D^{(n)}}{id}{kb} \right) \nonumber \\
	&-&\sum_j(1-P_{bc})\ep{D^{(n)}}{j}{b}\ep{S}{ia}{jc} \ , \\
  	 \ep{D^{(n+1)}}{ij}{ka}&=& \sum_{dl}(1-P_{ij})\left( -\ep{D^{(n)}}{il}{kd}\ep{S}{jd}{la}+\ep{S}{il}{kd}\ep{D^{(n)}}{jd}{la} \right) \nonumber \\
  	 &+& \sum_b (1-P_{ij})\ep{D^{(n)}}{j}{b}\ep{S}{ib}{ka}\ , \\
  	 \ep{D^{(n+1)}}{ia}{jb}&=& \sum_{kc}\left( -\ep{D^{(n)}}{ka}{jc}\ep{S}{ic}{kb}+\ep{S}{ka}{jc}\ep{D^{(n)}}{ic}{kb} \right)\nonumber\\
  	  &+&\sum_c \left(\ \ep{D^{(n)}}{a}{c}\ep{S}{ic}{jb} -\ep{D^{(n)}}{c}{b}\ep{S}{ia}{jc} \right)\nonumber \\
  	  &+&\sum_k \left( -\ep{D^{(n)}}{k}{j}\ep{S}{ia}{kb} +\ep{D^{(n)}}{i}{k}\ep{S}{ka}{jb} \right) \ , \\
  	  \ep{D^{(n+1)}}{ij}{ab}&=& -\sum_{kc}(1-P_{ij})(1-P_{ab})\ep{D^{(n)}}{ik}{ac}\ep{S}{jc}{kb} \ .
 \end{eqnarray}
 
Once decoupled, the effective interaction consists of three sectors. The first sector is the typical $pp$ interaction, for which the low-lying excited states of particles attached to the core can be calculated. The second sector is the hole-hole interaction, which can be used in hole-hole shell model calculations for particle removed systems. Another sector is the particle-hole channel, which can be used to calculate nuclei with protons removed and neutrons attached depending on the model space selected.

For hole-hole shell-model calculation, one can use standard shell-model codes. The expectation values of the one-body interaction are calculated via
 \begin{equation}
 	\langle ij | f_{pq}\{p^\dag q\} | kl \rangle =-f_{ki}\delta_{jl}-f_{lj}\delta_{ik}+f_{li}\delta_{jk}+f_{kj}\delta_{il} \ , 
 \end{equation}
 and two-body expectation values are
 \begin{equation}
 	\langle ij | f_{pqrs}\{p^\dag q^\dag rs\} | kl \rangle
 =f_{klij}  \ .
 \end{equation}
 The particle-hole interaction is more complicated. The one-body operator is applied as
  \begin{equation}
 	\langle ai^{-1} | f_{pq}\{p^\dag q\} | bj^{-1} \rangle =f_{ab}\delta_{ij}-f_{ji}\delta_{ab} \ .
 	\end{equation}
 The two-body operator needs an explicit Pandya transformation~\cite{Pandya1956}
 \begin{equation}
 	\langle ai^{-1}|V|bj^{-1}\rangle^{J}=-\sum_{J'}J'^2\begin{Bmatrix}
j_a & j_i & J\\
j_b & j_j & J'
\end{Bmatrix}  \langle aj|V|bi\rangle^{J'} \ .
\end{equation}
 
The interaction can also be re-normal-ordered with respect to a lighter core where all single-particle orbitals are particle states. The shell-model effective interaction from SMCC is non-Hermitian. To be used in standard shell model codes, the effective interaction can be made Hermitian through
\begin{equation}
	H^\text{hm}_\text{eff}=[\omega^\dag \omega]^{1/2}H^\text{nhm}_\text{eff}[\omega ^\dag \omega]^{-1/2} \ .
\end{equation}
Here, $\omega$ is the matrix that diagonalizes the non-Hermitian Hamiltonian $H^\text{nhm}_\text{eff}$.
 The Hermitian $H^\text{hm}_\text{eff}$ contains one and two-body terms and can be used in conventional shell-model calculations.
 
The truncation at the two-body level is a reasonable  first approximation in SMCC and VS-IMSRG. However, The off-diagonal three-body force $S_\text{3b}^\text{od}$ induced by the similarity transformation cannot be neglected for an accurate description of nuclei. The full three-body decoupling is computationally very expensive and challenging to implement. However, the leading  contribution from $S_\text{3b}^\text{od}$ is accessible through the perturbative approach and captures the important parts of three-body correlations~\cite{sun2018}. We discuss this correction next.
 
\begin{figure}[ht]
  \includegraphics[width=0.48\textwidth]{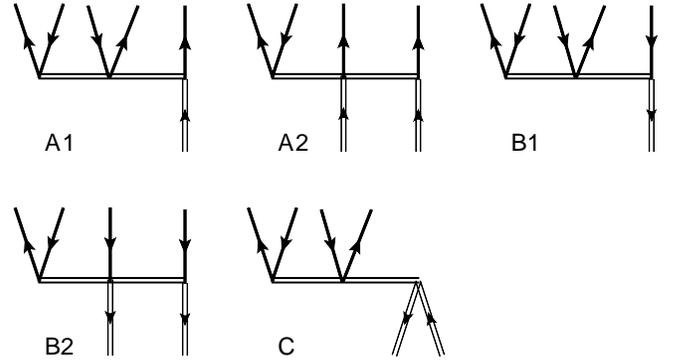}
  \caption{Diagrammatic representation of $S_\text{3b}$. The horizontal line is the $S_\text{3b}$ operator, with the particles indicated by incoming and outgoing arrow lines. The model-space particles and holes are indicated by incoming and outgoing double lines respectively. A1 and A2 are used in particle-particle decoupling, B1 and B2 for hole-hole decoupling, and all A, B, and C diagrams are used for particle-hole decoupling.}
\label{fig:3boff}
\end{figure}

Suppose we have decoupled the Hamiltonian at the two-body level, and the transformed Hamiltonian $\dbar{H}$ consists of one-body, two-body, and three-body terms, i.e. 
 \begin{equation}
	\dbar{H}=\dbar{H}_\text{1b}+\dbar{H}_\text{2b}+\dbar{H}_\text{3b} \ .
 \end{equation}
 Here, the leading induced three-body force is
  \begin{equation}
 	\dbar{H}_\text{3b}=[\sbar{H}_\text{2b},S_\text{2b}]_\text{3b} \ .
 \end{equation}
 Here we neglected induced three-body terms from the nested commutators. The operator $\dbar{H}_\text{3b}$ can be split into diagonal $\dbar{H}_\text{3b}^\text{da}$ and off-diagonal $\dbar{H}_\text{3b}^\text{od}$ contributions. Suppose $\dbar{H}_\text{3b}^\text{od}$ is driven to zero in a third similarity transformation 
 \begin{equation}\label{smcc3bdec}
 	Qe^{-S_\text{3b}}\dbar{H}e^{S_\text{3b}}P=0 \ ,
 \end{equation}
 where $S_\text{3b}$ couples the valence space and excluded space at the three-body level.
 This operator  would feed back to the two-body decoupling equation, breaking the two-body level decoupling condition $Q\dbar{H}P=0$. As an approximation, we assume this feedback can be neglected, because in a proper decoupling, the induced $\dbar{H}_\text{3b}$ should be small compared to $\dbar{H}_\text{2b}$. The left-hand side of Eq.~(\ref{smcc3bdec}) can therefore be written as
 \begin{eqnarray}
 	Qe^{-S_\text{3b}}\dbar{H}e^{S_\text{3b}}P&\approx &Q(1-S_\text{3b})\dbar{H}(1+S_\text{3b})P\\
 	&\approx &Q\dbar{H}P+Q[\dbar{H},S_\text{3b}]P
 	\label{smccdec3b2}.
 \end{eqnarray}
 Substitution of Eq.(\ref{smccdec3b2}) into  Eq.(\ref{smcc3bdec}) yields the decoupling equation that determine $S_\text{3b}$; these are
 \begin{eqnarray}
Q\dbar{H}_\text{1b}P+Q[\dbar{H},S_\text{3b}]_\text{1b}P&\approx&0\label{3bdec1}\ , \\
 Q\dbar{H}_\text{2b}P+Q[\dbar{H},S_\text{3b}]_\text{2b}P&\approx&0\label{3bdec2}\ , \\
 Q\dbar{H}_\text{3b}P+Q[\dbar{H},S_\text{3b}]_\text{3b}P&\approx&0\label{s3bdec} \ .
 \end{eqnarray}
 As we have assumed that $Q\dbar{H}P=0$ is preserved on the two-body level, Eqs.~(\ref{3bdec1}) and (\ref{3bdec2}) are trivial, whereas $S_\text{3b}$ can be determined by Eq.~(\ref{s3bdec}). If we keep only the diagonal one-body part in $\dbar{H}$, the operator $S_\text{3b}$ has the simple form
 \begin{equation}
 	\langle pqu |S_{3b}| rst\rangle =\frac{\langle pqu |\dbar{H}_{3b}| rst\rangle }{\dbar{H}_{pp}+\dbar{H}_{qq}+\dbar{H}_{uu}-\dbar{H}_{rr}-\dbar{H}_{ss}-\dbar{H}_{tt}}.
 \end{equation}
 Figure~\ref{fig:3boff} shows the diagrams included in $S_\text{3b}$. The diagrams A1 and A2 couple the valence particles to the excluded space. Diagrams B1 and B2 couple the valence holes and the excluded space, and diagram C couples the particle-hole states to the excluded space in the isospin $T_z=\pm1$ channels.
 
With $S_\text{3b}$, the effective shell model interaction becomes
  \begin{eqnarray}
H^\text{eff}_\text{1b}&=&P\dbar{H}_\text{1b}P+P[\dbar{H},S_\text{3b}]_\text{1b}P\label{v3b1} \ , \\
H^\text{eff}_\text{2b}&=& P\dbar{H}_\text{2b}P+P[\dbar{H},S_\text{3b}]_\text{2b}P
\label{v3b2} \ .
 \end{eqnarray}
The final effective interaction is thereby decoupled at the two-body level with perturbative three-body corrections included.

%% file: results.tex
\section{Results}
\label{sec:results}

We use  two  chiral interactions to compute the corresponding  shell-model effective interactions.
The first is the 1.8/2.0(EM) potential~\cite{hebeler2011} which results from a similarity-renormalization-group (SRG) transformation~\cite{bogner2007} at cutoff $\lambda=1.8$ fm$^{-1}$ of the N$^3$LO(EM) nucleon-nucleon potential from~\cite{entem2003}, and the three-body potential is given at N$^2$LO in chiral EFT~\cite{VanKolck1999,Epelbaum2002,Hebeler2015a} with a cutoff $\Lambda_{NNN}=2.0$~fm$^{-1}$. The second potential is the $\Delta \text{NNLO}_\text{GO}$ with a cutoff $\Lambda=394$~MeV~\cite{jiang2020}. This nucleon-nucleon and three-nucleon interaction includes $\Delta$ degrees of freedom. With both interactions we perform a Hartree-Fock (HF) calculation in the harmonic oscillator basis with a frequency of $\hbar\omega=16$~MeV. The model space is spanned by 13 shells ($N_{\rm{max}}=12$), and the three-body matrix elements are truncated at $E_{\rm{3max}}=16\hbar\omega$. The two-body and three-body interaction is then normal ordered with respect to the $^{100}$Sn core. The CC calculations are performed with the normal-ordered two-body interaction.

We use $^{100}$Sn as the core, with the model space spanned by neutron-particle and proton-hole states. Using the effective shell-model interactions, we studied the nuclei  south-east of $^{100}$Sn by removing protons and attaching neutrons. To decouple the core, we employed the CCSDT-1 approximation~\cite{lee1984} for both potentials.
Using the highly optimized nuclear tensor contraction library (NTCL)~\cite{ntcl2020} we are able to perform these calculations in the full space (without truncating the number of triples amplitudes) on {\sc Summit}, the supercomputer of the U.S. Department of Energy with a peak performance of 200 petaflops which is operated by the Oak Ridge Leadership Computing Facility at Oak Ridge National Laboratory.

The resulting binding energy of $^{100}$Sn is $-837$ MeV and $-816$ MeV for 1.8/2.0(EM) and $\Delta \text{NNLO}_\text{GO}$, respectively, compared to the datum of $-825.3$~MeV. The proton model space includes the $p_{3/2}$, $p_{1/2}$, and $g_{9/2}$ orbitals, and the neutron space consists of  $g_{7/2}$, $d_{5/2}$, $d_{3/2}$, $s_{1/2}$, and $h_{11/2}$. We decouple the Hamiltonian at the two-body level, and include the off-diagonal three-body terms perturbatively.

\begin{table}[ht]
\begin{tabular}{c|r|r|r}
\hline\hline
\multicolumn{2}{c|}{s.p. Energy (MeV}  & 1.8/2.0(EM) & $\Delta\text{NNLO}_\text{GO}$ \\ \hline
\multirow{3}{*}{holes}     & $p_{3/2}$  & $-6.502 $         & $-4.798  $                           \\ \cline{2-4} 
                           & $p_{1/2}$  & $-4.871 $         & $-3.159  $                          \\ \cline{2-4} 
                           & $g_{9/2}$  & $-3.106 $          & $-2.267 $                          \\ \hline
\multirow{5}{*}{particles} & $g_{7/2}$  & $-10.832$          & $-10.072$                            \\ \cline{2-4} 
                           & $d_{5/2}$  & $-10.548$          & $-9.375 $                           \\ \cline{2-4} 
                           & $d_{3/2}$  & $-8.054 $         & $-6.811  $                           \\ \cline{2-4} 
                           & $s_{1/2}$  & $-7.789 $         & $-6.220  $                           \\ \cline{2-4} 
                           & $h_{11/2}$ & $-5.596 $          & $-5.012$                           \\ \hline\hline
\end{tabular}
\caption{Single-particle energies calculated with the 1.8/2.0(EM) and $\Delta \text{NNLO}_\text{GO}$ potentials. $p_{3/2}$, $p_{1/2}$ and $g_{9/2}$ are hole states with respect to $^{100}$Sn, and the remaining states are particle states}
\label{table1}
\end{table}

Table~\ref{table1} shows the calculated single-particle and single-hole energies from the two chiral potentials. For $^{101}$Sn, the experimental splitting is a 171~keV between the $5/2^+$ and the $7/2^+$ states~\cite{seweryniak2006,darby2010}. In our calculation the 1.8/2.0(EM) yields the $E(g_{7/2})$ 284~keV lower than $E(d_{5/2})$, which agrees with the PA-EOM-CC and VS-IM-SRG calculations~\cite{morris2018}. Meanwhile, the splitting from $\Delta \text{NNLO}_\text{GO}$ is 700~keV, which is a bit larger than the datum. Both potentials predict a $9/2^+$ ground state of $^{99}$In, which has a single-hole configuration. Matrix elements of these interactions are available in the Supplemental Material. 

Figure~\ref{2bspec} shows the calculated binding energies of two-particle ($^{102}$Sn), two-hole ($^{98}$Cd) and one-particle-one-hole ($^{100}$In) nuclei with respect to the ground state of $^{100}$Sn. The 1.8/2.0(EM) potential is in  agreement with the experimental data. The $\Delta \text{NNLO}_\text{GO}$ somewhat underestimates $^{102}$Sn and overbinds the two-proton system $^{98}$Cd. The two potentials coincide in $^{100}$In. In a simple non-interacting shell-model picture, the ground state of $^{100}$In depends on $E(\pi g_{7/2})-E(\nu g_{9/2}$), and the coincidence in $^{100}$In is indicated from our calculated single-particle energies.

\begin{figure}[ht]
  \includegraphics[width=0.5\textwidth]{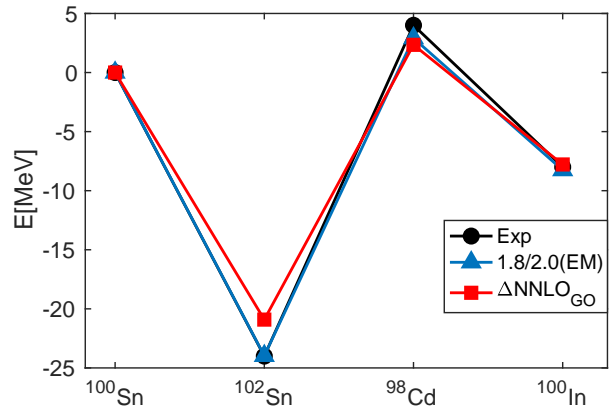}
  \caption{Ground-state energies of $^{102}$Sn, $^{98}$Cd, and $^{100}$In relative to $^{100}$Sn, calculated with 1.8/2.0(EM) and $\Delta \text{NNLO}_\text{GO}$ potential and compared to experimental data}
  \label{2bspec}
\end{figure}

\begin{figure}[ht]
  \includegraphics[width=\linewidth]{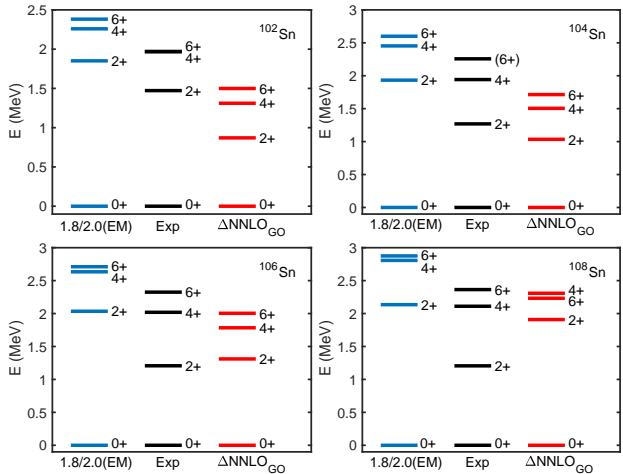}
  \caption{Calculated low-lying excited states in Sn isotopes with the 1.8/2.0(EM) (blue) and $\Delta \text{NNLO}_\text{GO}$ (red) potentials, and compared to data (black).}
  \label{sniso}
\end{figure}

The tin isotopes depend only on the neutron-neutron interaction, i.e. the particle-particle interaction constructed in this work. Figure~\ref{sniso} shows the low-lying excited states calculated for ${^{102-104}}$Sn. 
The 1.8/2.0(EM) potential yields a near constant $2^+$ energy, in agreement with the VS-IMSRG calculation of Ref.~\cite{morris2018}. The $\Delta \text{NNLO}_\text{GO}$ potential also gives a near constant $E(2^+$) energy (with the exception of $^{108}$Sn), but generally lower in excitation energy than obtained using the 1.8/2.0(EM) potential. Data falls in-between the two employed potentials for $^{102,104}$Sn, for $^{106}$Sn the results obtained with $\Delta \text{NNLO}_\text{GO}$ agrees well with data, while for $^{108}$Sn both potentials overestimate the $E(2^+)$ energy.  The difference in the $E(2^+)$ energies obtained from the two potentials stems mostly from the associated single-particle energies given in Table~\ref{table1}.

The cadmium isotopes are obtained by removing two protons from $^{100}$Sn core and adding neutrons. Our calculations indicate that the removed neutrons are mainly from $g_{9/2}$ and the spectra of cadmium isotopes are mostly determined by the neutron configurations. Similar to the tin isotopes, the $2^+$ states for even-even cadmiums from 1.8/2.0(EM) are higher than the data, and the $\Delta \text{NNLO}_\text{GO}$ shows a slight increase of $E(2^+)$ as the neutron number increases. 

\begin{figure}[ht]
  \includegraphics[width=\linewidth]{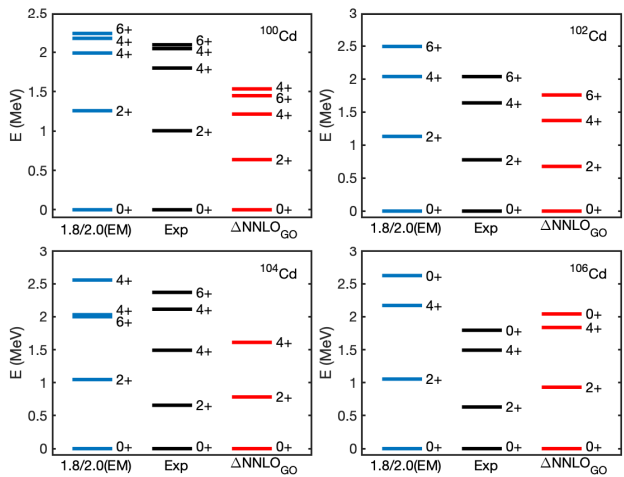}
  \caption{Calculated low-lying excited states in Cd isotopes with the 1.8/2.0(EM) (blue) and $\Delta \text{NNLO}_\text{GO}$ (red) potentials, and compared to data (black).}
\end{figure}

The two potentials produce more differences in the indium isotopes, which remove one proton and add neutrons to the $^{100}$Sn core. The odd-mass indium nuclei have $9/2^+$ ground states and a first $1/2^-$ low-lying excited state (with the exception of $^{101}$In) according to the experimental data. The 1.8/2.0(EM) potential reproduces the correct order of the ground states and the $1/2^-$ excited state in $^{101-107}$In. The computed $1/2^-$ state is generally  higher than the data, which may be due to a too large gap between the $g_{9/2}$ and $p_{1/2}$ orbitals. The $\Delta \text{NNLO}_\text{GO}$  reproduces the correct level ordering in $^{101,103}$In, but fails to reproduce the correct ground state of $^{105,107}$In. In contrast to tin and cadmium isotopes, the indium isotopes exhibit more proton and neutron correlations. Finally, we note that uncertainties stemming from model-space truncations and the approximate treatment of normal-ordered three-body forces away from the $^{100}$Sn core will impact the results. In the recent work~\cite{mougeot2021} we estimated that empirical pairing gaps in this mass region computed with the SMCC and employing the same Hamiltonians carried an uncertainty of about $\pm 0.2$~MeV.

\begin{figure}[ht]
  \includegraphics[width=\linewidth]{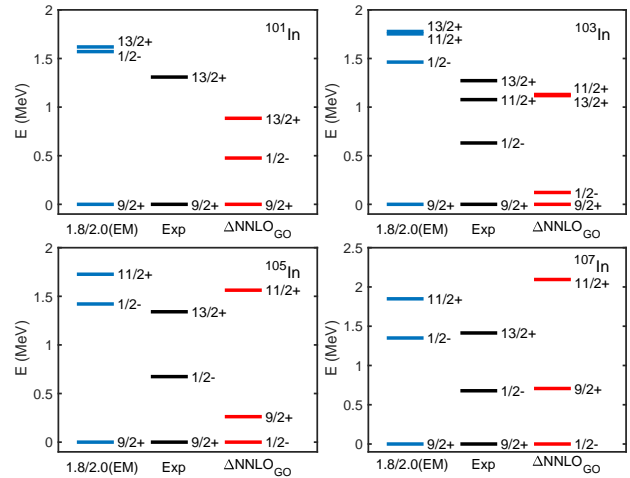}
  \caption{Calculated low-lying excited states in In isotopes with the 1.8/2.0(EM) (blue) and $\Delta \text{NNLO}_\text{GO}$ (red) potentials, and compared to data (black).}
\end{figure}

\section{Summary}
We presented a systematic derivation of the particle-hole variant of the shell-model coupled-cluster method to compute nuclei in the vicinity of $^{100}$Sn. The shell-model effective interaction is defined in a model space consisting of both particles and holes. The decoupling of the model space from the excluded space is accomplished at the two-body level, and the induced off-diagonal three-body terms are included perturbatively. For nuclei in the vicinity of $^{100}$Sn the particle-hole effective interaction benefits from a more favorable reference state and more realistic mean-field than, e.g., taking a $^{88}$Sr core. The computational resources required for nuclei close to the core are also reduced by introducing explicit hole states in the shell model. The method is validated through cross-benchmark with the existing ab-into methods. We derived effective interactions from two sets of chiral potentials that include nucleon-nucleon and three-nucleon forces, and computed properties of tin, indium, and cadmium isotopes. Binding energies and spectra exhibit a good to fair agreement with data.